\documentclass[runningheads]{llncs}
\usepackage{graphicx}
\usepackage{xcolor}
\usepackage{xurl}
\usepackage{comment}

\PassOptionsToPackage{hyphens}{url}\usepackage{hyperref}

\usepackage{pgfplots}
\pgfplotsset{compat=newest}

\begin{document}
\title{Minimally Intrusive Access Management to Content Delivery Networks based on Performance Models and Access Patterns}  

%
\titlerunning{Minimally Intrusive Access Management to CDNs}
%
\author{Lenise M. V. Rodrigues\inst{1,}\inst{3}\and Daniel Sadoc Menasché\inst{1} \and \\ Arthur Serra\inst{3} \and Antonio A. de Aragão Rocha\inst{2}}
%
\authorrunning{L. Rodrigues et al.}
%
\institute{Federal University of Rio de Janeiro (UFRJ), Brazil  \\ 
\and
Fluminense Federal University (UFF), Rio de Janeiro, Brazil  \\ 
\and
Globo, Rio de Janeiro, Brazil\\
}
\maketitle              
\begin{abstract}
  This paper presents an approach to managing access to Content Delivery Networks (CDNs), focusing on combating the misuse of tokens through performance analysis and statistical access patterns. In particular, we explore the impact of token sharing on the content delivery infrastructure, proposing the definition of acceptable request limits to detect and block abnormal accesses. Additionally, we introduce countermeasures against piracy, such as degrading the quality of service for pirate users to discourage them from illegal sharing, and using queuing models to quantify system performance in different piracy scenarios. Adopting these measures can improve the consistency and efficiency of CDN access and cost management, protecting the infrastructure and the legitimate user experience.

\end{abstract}
\section{Introduction}
\label{sec:introduction}
Over the past decade, there has been a tremendous growth of Internet content streaming platforms. As an example, 43.4\% of Brazilian households with TVs now use paid streaming service.\footnote{\url{https://agenciadenoticias.ibge.gov.br/agencia-noticias/2012-agencia-de-noticias/noticias/38306-em-2022-streaming-estava-presente-em-43-4-dos-domicilios-com-tv} (last accessed 29 August, 2024)} Managing access to determine which users have rights to access specific content is a challenge and the central theme of this work. In particular, our focus is on studying the impacts of illegal access token sharing to view live content using the infrastructure of a national Brazilian Content Delivery Network (CDN) that serves millions of users daily.

The CDN aims to estimate the costs associated with non-legitimate content consumption and to block the corresponding users without impacting legitimate users due to classification errors. In particular, when legitimate users are classified as suspicious, they may face a denial of service due to a false positive. This motivates a thorough analysis before revoking tokens. However, thorough analysis is expensive, consuming resources such as computational power and energy, and causing delays, which can affect legitimate users 
and can preclude the timely blocking of illegal users~\cite{10.17487/RFC9449,9833728}.

Given the above challenges, we pose the following two key research questions:
\begin{itemize}
\item \textbf{What is the performance loss of a CDN when serving illegitimate content?} In particular, how does such a loss increase the costs of the CDN and/or affect the quality of service (QoS) for users?
\item As the implemented solutions need to be simple, \textbf{how do the simplest access management solutions behave in real networks?}
\end{itemize}
To answer the first question, we turn to queuing models (Section~\ref{sec:model}). In particular, we consider models like $M/M/1$, $M/M/1$ with burst arrivals (to capture malicious users), and $M/M/1$ with priorities (to capture   strategies that favor legitimate users). Through such models, we illustrate some elements involved in the performance of access management systems. To answer the second question, we analyzed data from a national CDN, which serves one of the largest Brazilian content producers, with millions of daily hits (Section~\ref{sec:evaluation}). We show that access management policies based on thresholds already bring significant results in mitigating the actions of malicious users, with few false positives.

In summary, our contribution is twofold. 

\textbf{Queuing models for performance analysis. } We propose queuing models that allow us to understand how system performance varies depending on the level of piracy. In particular, we analyze both the case in which no countermeasures are being put in place (Section~\ref{sec:problem_model}) and the case in which such countermeasures are implemented (Section~\ref{sec:mitigation_model}). 

\textbf{Real data analysis. } Then, we analyze real data on access management (Section~\ref{sec:evaluation}). The data suggests that even simple policies, based on thresholds, are sufficient to detect a significant fraction of inappropriate access. 

\textbf{Outline. } The remainder of this paper is organized as follows. 
  Section~\ref{sec:related-work} presents related literature. Section~\ref{sec:model} introduces models and Section~\ref{sec:evaluation} measurements. Finally, Section~\ref{sec:conclusion} concludes.

\section{Related Work}
\label{sec:related-work}
Below, we present some research lines related to the main themes covered in this article. 

\textbf{Traffic characterization to CDNs. } The security of CDN networks is the subject of numerous works, for example, focusing on statistical aspects related to anomaly detection or practical issues of system implementation, such as Akamai~\cite{gillman2015protecting,gonccalves2020model}. Traffic characterization is an important step towards increasing network security. Once “normal” traffic is characterized, one can also attempt to characterize anomalous traffic. The advantage of having the characterization of both consists of greater accuracy in detecting anomalies. The disadvantage is that one may not detect anomalies if their signatures have never been observed before. In this work, we use simple threshold-based methods to determine whether certain accesses are authentic or not.

\textbf{Piracy. } The piracy problem in Brazil was studied, for example, in~\cite{dent2020digital}. 
This is a multifaceted problem, involving legal, human, and technological aspects. 
In this work, we focus on technological aspects. In particular, we focus on system performance evaluation and monitoring of real systems. 
In future work, we also intend to evaluate economic aspects, for example, related to costs of pirated equipment. 
A preliminary, theoretical analysis involving prices is presented in Section~\ref{sec:mitigation_model}.

\textbf{Piracy in Streaming} The issue of piracy in the streaming sector has recently gained significant attention, with content providers prioritizing strategies to reduce financial losses. In~\cite{10.1145/3658664.3659663}, the discussion revolves around the various strategies by video pirates to bypass protections. One of the strategies is the token sharing, which is a central theme of this work. In~\cite{10.1145/3638036.3640276}, the authors introduce token mechanisms to prevent unauthorized access and misuse of CDN resources. Those mechanisms complement the ones introduced in the remainder of this paper. 


\textbf{Queuing models for detecting misuse of networks.} In the context of covert communication, many works are dedicated to understanding the effect of misuse of resources~\cite{jiang2021covert}. 
In these cases, authors focus on the problem of resource theft without owners realizing the problem. In this article, on the other hand, we primarily consider the effects of resource theft on legitimate users, who have their quality of service affected. 
The problem of detecting misuse of resources was considered, for example, in~\cite{rufino2020improving}. 
However, while in~\cite{rufino2020improving} and~\cite{reza2022fundamental} the problem of denial of service was considered, in the present work we consider that pirates will affect the network, but not to the extent of generating denial of service, which would end up being undesirable for pirates as well.


\section{Models}
\label{sec:model}
In this section, we consider models to capture system performance before and after adopting countermeasures. 

\begin{figure}[t] 
\centering 
\includegraphics[width=\textwidth]{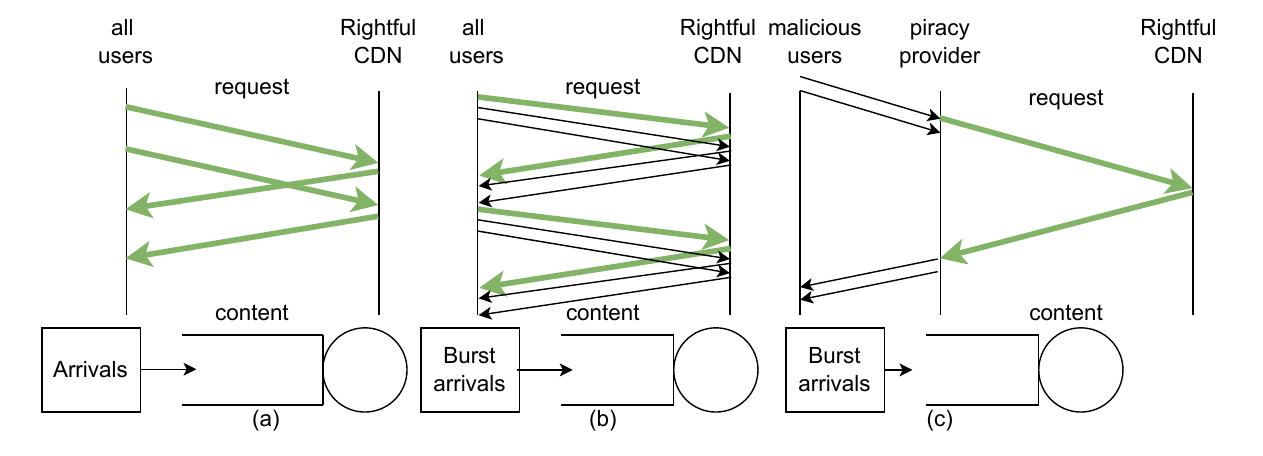} \caption{Illustration of the 3 scenarios considered: (a) standard operation; (b) hackers using the CDN for pirated content replication; (c) hackers using their infrastructure to replicate content} \label{fig:3cenarios} 
\end{figure}

\subsection{Problem setup and queuing models} 
\label{sec:problem_model} 
Without countermeasures, we face the scenarios illustrated in Figure~\ref{fig:3cenarios}. 
In the initial scenario (Figure~\ref{fig:3cenarios}(a)) we have legitimate requests being sent to the CDN that responds with content. 
In the intermediate scenario (Figure~\ref{fig:3cenarios}(b)), we have malicious users via piracy providers generating requests for content that is served by the original content provider's own CDN. 
Finally, in the last scenario (Figure~\ref{fig:3cenarios}(c)), we have malicious users via piracy providers generating requests that are served by an infrastructure maintained by those piracy providers. 

To capture the 3 scenarios, we use queuing models. 
In this work, we consider the simplest models, aiming to quantify the effects of the three types of pirated content sharing. 
Therefore, in the original scenario, we considered an M/M/1 queue. 
In the intermediate scenario, we consider an M/M/1 queue with arrivals in bursts (\emph{batch arrivals}), assuming that each request corresponds to the service of $b$ users (for example, $b$ pirate users). 
Finally, in the last case, we assume that each request also corresponds to the service of $b$ pirate users. 
However, if some legitimate users switch to the pirate strategy, this may reduce the burden on the legitimate CDN. 
To capture this phenomenon, we assume that the arrival rate now becomes $\lambda'$. 
Unless otherwise stated, we assume that $\lambda'=\lambda/b$. 
Therefore, the arrival rate of requests to the CDN in the scenario in Figure~\ref{fig:3cenarios}(c) is equal to $\lambda/b$, but the service rate to the users, taking into account the hacker infrastructure, is equal to $\lambda=b \cdot \lambda/b$.

Note that the scenarios in Figures~\ref{fig:3cenarios}(a) and~\ref{fig:3cenarios}(c) are distinguished only by the fact that in the first the arrivals follow a Poisson flow, while in the last they follow a burst flow. 
Although this model corresponds to a specific instantiation of the problem in question, in which the total rate of users served per unit of time remains constant between the scenarios in Figures~\ref{fig:3cenarios}(a) and \ref{fig:3cenarios}(c), we believe that it already serves as a first step to illustrate the effect of hackers on system performance, taking into account both the legitimate CDN network and the hackers' infrastructure.

Assume that the bursts arrive according to a Poisson flow with rate $\lambda$, and each service of each request takes exponentially distributed time with mean $1/\mu$. 
The average waiting time in an M/M/1 queue with burst arrivals of size $b$ is given by~\cite{ghimire2014mathematical}, $ 
    T=({1+b})/({2 \mu(1-\rho)})$ 
where $\rho=\lambda/\mu$. 
In the three considered scenarios, the waiting time for customers will be given by the following expressions, respectively, \begin{equation} 
    T_{orig}=\frac{1}{ \mu(1-\rho)}, \quad T_{CDN}=\frac{1+b}{2 \mu(1-\rho)}, \quad T_ {infra}=\frac{1+b}{2 \mu(1-\lambda/(b\mu ))}. 
\end{equation}

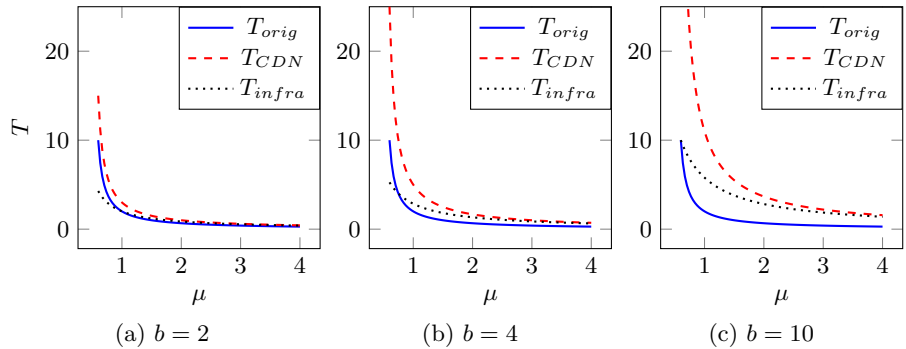
\begin{figure} \centering
    \begin{tabular}{ccc} 
\begin{tikzpicture}
    \begin{axis}[  width=4.8cm, 
        height=4.8cm, 
        xlabel={$\mu$},
        ylabel={$T$},
        domain=0.6:4, ymax=25, 
        samples=100,
        legend style={at={(1,1)},anchor=north east},
    ]
    \addplot[blue, thick] {1 / ( x * (1 - 0.5/x))}; 
    \addlegendentry{$T_{orig}$}

    \addplot[red, thick, dashed] {(1 + 2) / (2 * x * (1 - 0.5/x))}; 
    \addlegendentry{$T_{CDN}$}

    \addplot[black, thick, dotted] {(1 + 2) / (2 * x * (1 - 0.5/(2 * x)))}; 
    \addlegendentry{$T_{infra}$}
    
    \end{axis}
\end{tikzpicture}
         & \begin{tikzpicture}
    \begin{axis}[  width=4.8cm, 
        height=4.8cm, 
        xlabel={$\mu$},
         domain=0.6:4, ymax=25, 
        samples=100,
        legend style={at={(1,1)},anchor=north east},
    ]
    \addplot[blue, thick] {1 / ( x * (1 - 0.5/x))}; 
    \addlegendentry{$T_{orig}$}

    \addplot[red, thick, dashed] {(1 + 4) / (2 * x * (1 - 0.5/x))}; 
    \addlegendentry{$T_{CDN}$}

    \addplot[black, thick, dotted] {(1 + 4) / (2 * x * (1 - 0.5/(4 * x)))}; 
    \addlegendentry{$T_{infra}$}
    
    \end{axis}
\end{tikzpicture}   & \begin{tikzpicture}
    \begin{axis}[  width=4.8cm, 
        height=4.8cm, 
        xlabel={$\mu$},
         domain=0.6:4, ymax=25, 
        samples=100,
        legend style={at={(1,1)},anchor=north east},
    ]
    \addplot[blue, thick] {1 / ( x * (1 - 0.5/x))}; 
    \addlegendentry{$T_{orig}$}

    \addplot[red, thick, dashed] {(1 + 10) / (2 * x * (1 - 0.5/x))}; 
    \addlegendentry{$T_{CDN}$}

    \addplot[black, thick, dotted] {(1 + 10) / (2 * x * (1 - 0.5/(10 * x)))}; 
    \addlegendentry{$T_{infra}$}
    
    \end{axis}
\end{tikzpicture}  \\
(a) $b=2$ & (b) $b=4$ & (c) $b=10$
    \end{tabular}
    \caption{Waiting time experienced by clients in the case of a system with 2, 4, or 10 pirate clients for each legal request.}  \label{fig:tempoespera}
\end{figure}

Figure~\ref{fig:tempoespera} illustrates the behavior of the model. 
In Figure~\ref{fig:tempoespera} we consider $\lambda=0.5$, $\mu=0.6$ and $b=1$ as our reference scenario. 
The three systems illustrated in this figure describe the three scenarios listed above, represented by the full (blue), hatched (red), and dotted (black) curves, respectively. 
As the system's service capacity increases, that is, as $\mu$ grows between 0.6 and 4, the difference between the systems' performance decreases. In particular, when $b=2$ or $b=4$, the difference between the systems when $\mu=4$ is practically imperceptible. 
If the system is over-provisioned, even with the presence of pirates, the system performance remains practically unchanged. 
Now consider the increase in the number of pirate users per legitimate request. 
When $b$ increases between 2, 4, and 10, the difference between the red, blue, and black curves increases. 
This corresponds to the fact that the greater the number of pirate users, the greater the load on the CDN or pirate infrastructure. 
It is also worth noting that according to the model considered, performance with pirated infrastructure is generally between the performance of the original system and the performance of the system that makes exclusive use of the CDN. 
The model captures the intuition that pirated infrastructure will compensate for overloads on the target CDN.

\subsection{Problem mitigation} 
\label{sec:mitigation_model} 
One of the simplest countermeasures against piracy is to degrade the quality of service of pirate users, to such an extent that it is not worth paying for the pirated service. 
If it is better to use the original service, rather than the pirated one, there will be no reason to use the pirated service. 
Degrading QoS is a milder and gentler countermeasure against piracy than simply banning any pirate accounts, especially taking into account false positives.

Let $p$ be the price of the service and $T$ be the average service time for each request. 
Let $C$ be the user cost, which depends on QoS and price, with the cost associated with QoS being given by the service time, $$C=\alpha T+p,$$ where $\alpha$ is a constant that balances between the weight of price and QoS. 
The idea of the model is to capture the fact that the pirate system has lower $p$ but higher $T$. 
Let $p_i$ and $T_i$ be the price and service time of the illegal system, and $p_l$ and $T_l$ be the price and service time of the legal system. 
If the following condition is met, it is worth using the legal system, \begin{equation} 
    C_l \leq C_i \Rightarrow \alpha T_l + p_l \leq \alpha T_i + p_i.
\end{equation}
Assuming that the illegal price is a fraction $\beta$ of the legal price, $0 \leq \beta \leq 1$, $p_i = \beta \cdot p_l$, the condition is given by
\begin{equation} 
    p_l \leq \alpha \frac{T_i-T_l}{1-\beta}=P_l.
\end{equation}
We refer to the quantity on the right, $P_l$, as the limiting legal price. If the legal price is below this threshold, it is worth using the legal service.

In the simplest case, we have two queues, with the priority queue serving requests with priority, and without preemption, from customers classified as authentic. 
In this case, assuming all exponentially distributed services, we have~\cite{harchol2013performance}, 
\begin{equation} 
    T_i =\frac{\rho/\mu}{(1-\rho_l)(1-\rho_l-\rho_i)}, \qquad T_l = \frac{\rho/\mu}{1-\rho_l} 
\end{equation} 
where $\rho=\rho_l+\rho_i$, $\rho_l=q \lambda/\mu$, $\rho_i = (1-q) \lambda/\mu$, and $q$ is the fraction of services corresponding to legal customers.

\begin{figure}
    \centering \vspace{-0.1in}
    \begin{tabular}{cc} 
    \begin{tikzpicture}
    \centering
        \begin{axis}[
            width=5cm, 
            height=5cm, 
            xlabel={$q$},
            ylabel={$P_l$},
            domain=0:0.3,
            samples=100,
            ymax=2,
            ymin=0, 
            legend style={at={(1,0.5)},anchor=north east} 
        ]
        \pgfmathsetmacro{\rho}{0.5}
            \pgfmathsetmacro{\lambda}{0.5}
        \pgfmathsetmacro{\mu}{1}
        \pgfmathsetmacro{\rhol}{  \lambda}
        \pgfmathsetmacro{\rhoi}{\lambda}
        \pgfmathsetmacro{\alpha}{2}
        
        \pgfmathsetmacro{\betaA}{0.2}
        \pgfmathsetmacro{\betaB}{0.3}
        \pgfmathsetmacro{\betaC}{0.4}
        
        \addplot[blue, thick] {\alpha * (((\lambda / \mu) / ((1 - x*\lambda/\mu) * (1 - x*\lambda/\mu - (1-x)*\lambda/\mu))) - ((\lambda / \mu) / (1 - x*\lambda/\mu))) / (1 - \betaA)};
        \addlegendentry{$\beta=0.2$}
        
        \addplot[red, thick, dashed] {\alpha * (((\lambda / \mu) / ((1 - x*\lambda/\mu) * (1 - x*\lambda/\mu - (1-x)*\lambda/\mu))) - ((\lambda / \mu) / (1 - x*\lambda/\mu))) / (1 - \betaB)};
        \addlegendentry{$\beta=0.3$}
        
        \addplot[black, thick, dotted] {\alpha * (((\lambda / \mu) / ((1 - x*\lambda/\mu) * (1 - x*\lambda/\mu - (1-x)*\lambda/\mu))) - ((\lambda / \mu) / (1 - x*\lambda/\mu))) / (1 - \betaC)};
        \addlegendentry{$\beta=0.4$}
        
        \end{axis}
    \end{tikzpicture} &
    
    \begin{tikzpicture}
    \centering
        \begin{axis}[
            width=5cm,
            height=5cm,
            xlabel={$\mu$},
             domain=0.5:2,
            samples=100,
            ymax=2,
            ymin=0,
            legend style={at={(1,1)},anchor=north east} 
        ]
        \pgfmathsetmacro{\rho}{0.5}
            \pgfmathsetmacro{\lambda}{0.5}
        \pgfmathsetmacro{\rhol}{  \lambda}
        \pgfmathsetmacro{\rhoi}{\lambda}
        \pgfmathsetmacro{\alpha}{2}
        \pgfmathsetmacro{\betaA}{0.2}
                \pgfmathsetmacro{\betaB}{0.3}
                \pgfmathsetmacro{\betaC}{0.4}

        \pgfmathsetmacro{\qA}{0.1}
        \pgfmathsetmacro{\qB}{0.1}
         
        \addplot[blue, thick] {\alpha * (((\lambda / x) / ((1 - \qA*\lambda/x) * (1 - \qA*\lambda/x - (1-\qA)*\lambda/x))) - ((\lambda / x) / (1 - \qA*\lambda/x))) / (1 - \betaA)};
        \addlegendentry{$\beta=0.2$}

        \addplot[red, thick, dashed] {\alpha * (((\lambda / x) / ((1 - \qB*\lambda/x) * (1 - \qB*\lambda/x - (1-\qB)*\lambda/x))) - ((\lambda / x) / (1 - \qB*\lambda/x))) / (1 - \betaB)};
        \addlegendentry{$\beta=0.3$}

     \addplot[black, thick, dotted] {\alpha * (((\lambda / x) / ((1 - \qB*\lambda/x) * (1 - \qB*\lambda/x - (1-\qB)*\lambda/x))) - ((\lambda / x) / (1 - \qB*\lambda/x))) / (1 - \betaC)};
        \addlegendentry{$\beta=0.4$}

        \end{axis}
    \end{tikzpicture}  \\
    (a) & (b) 
    \end{tabular}
    \caption{Legal price threshold $P_l$ below which the legal service is worthwhile}\label{fig:threshold} \vspace{-0.1in}
\end{figure}
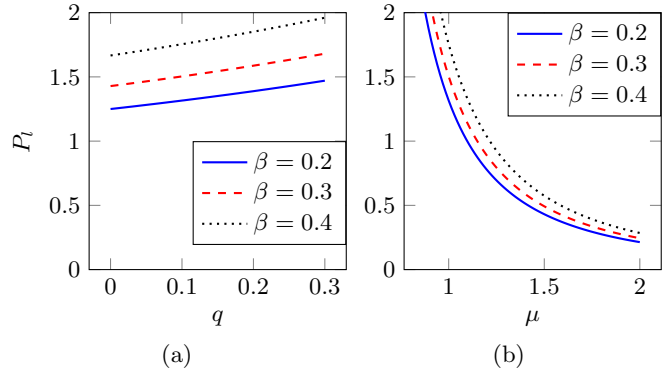

Figure~\ref{fig:threshold} illustrates the legal price threshold $P_l$ below which it is worth using the legal service (Figure~\ref{fig:threshold}(a) varying $q$ and Fig.~\ref{fig:threshold}(b) varying $\mu$). 
To generate the curve, we use $\rho=\lambda=0.5$, $\mu=1$ and $\alpha=2$. We note that as the fraction of legal users, $q$, increases, the overhead on the illegal users increases and, according to the proposed model, the QoS of pirate users degrades more than the QoS of legal users. 
Thus, the minimum price below which it is worth paying for the legitimate service increases. 
Alternatively, as $\beta$ increases, the price threshold also increases. 
After all, if the pirated service becomes more expensive, the minimum amount below which it is worth paying for the original service also increases. Finally, as $\mu$ increases, Figure~\ref{fig:threshold}(b) shows that users have less incentive to use the legal service, as the illegal alternative will already provide similar QoS.

\section{Evaluation}
\label{sec:evaluation}
In this section, we illustrate simple access control strategies using data from a national CDN that serves millions of users. We are focusing on the scenario shown in Figure~\ref{fig:3cenarios}(b), where we aim to block users who are suspected of misusing the legitimate CDN infrastructure. To achieve this, we use two approaches. First, we analyze the number of unique IPs associated with each user (Section~\ref{sec:volumetry1}). Users linked to many IPs are flagged as suspicious. However, it's important to note that multiple IPs can be legitimate, for example, when a user switches networks. To address this, we also examine the number of requests made by each user (Section~\ref{sec:volumetry2}). A summary of the considered approaches is presented in Table~\ref{tab:my_label}.

\subsection{How to evaluate abusive consumption in practice?} \label{sec:impacto-cdn} 
Next, we present some relevant observations from a practitioner's standpoint on how to evaluate abusive consumption in practice:
\begin{itemize} 
    \item Sessions that share tokens are not distinguishable through traditional monitoring tools, i.e., users running different streaming sessions but using the same token may not be distinguishable from the standpoint of certain monitoring tools. 
    
    \item Sessions do not require additional authentication from a central entity to play videos once they have access to tokens, i.e., once a user has a URL with an access token, the user can play the video from anywhere, as long as this token is valid. 
    \item  A video server can be affected by a single, indiscriminately distributed token. 
    \end{itemize}

    
    The video delivery process is summarized as follows: \begin{enumerate} 
        \item A user presses a button in the video player to start watching content. 
        \item The video player requests, through an API, a URL for that content, containing a token for authentication. If the user has the right to watch that content, the user should be able to receive the URL with the token. 
        \item In possession of this URL, the video player issues multiple requests for small chunks of video. 
    \end{enumerate} 
 
In the considered pirate token-sharing scenario of Figure~\ref{fig:3cenarios}(b), for each legitimate user the above steps are repeated $b$ times, one time per pirate user behind each legitimate user. All $b$ requests will leverage the same \emph{access token} but will correspond to different destinations.


\begin{table}[t] 
    \centering 
    \caption{Approaches for blocking tokens, remembering that each token is associated with a session, and is contained within a URL} 
    \begin{tabular}{c|c|c|c} & Metric adopted & Blocking & Problem \\ 
    \hline \hline 
    Approach 1 & IPs per user & tokens & there are still (few) false positives left\\ 
    \hline 
    Approach 2 & requests by IP by  user & tokens & to be determined  \end{tabular} 
    \label{tab:my_label}
\end{table}

\begin{figure}[t]
    \centering
    \begin{tabular}{c}
\includegraphics[width=1\textwidth]{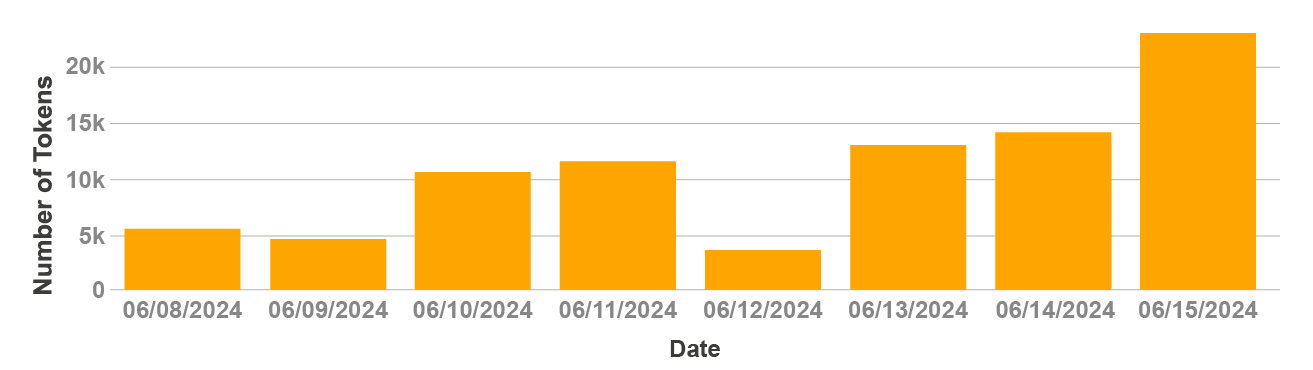}  \\
(a) \\
\includegraphics[width=1\textwidth]{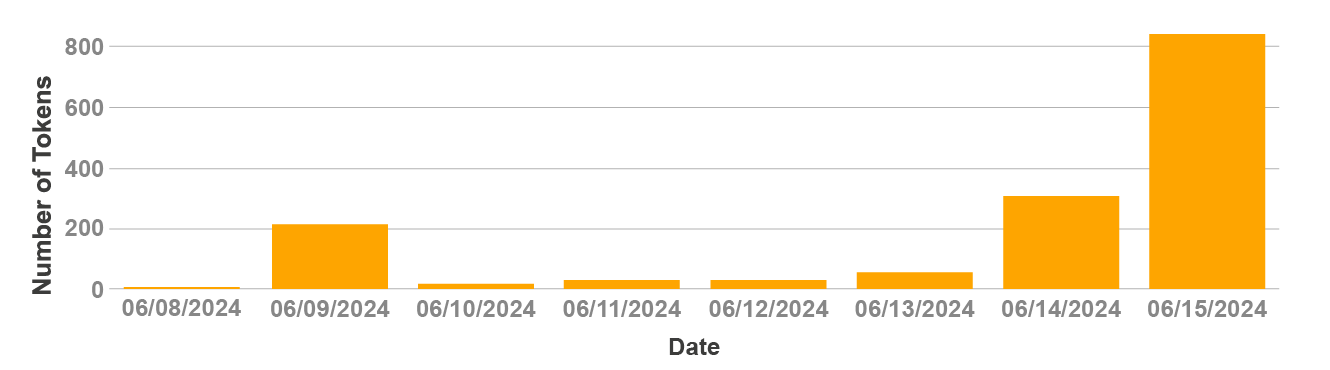} \\
(b) \end{tabular}
    \caption{Suspicious tokens per day (a) all cases and (b) non-recurring cases.}
    \label{fig:blocks-by-day}
\end{figure}

\subsection{Analysis of suspicious consumption occurrences via volumetry: IPs per user} 
\label{sec:volumetry1} 
A token is considered suspicious if it violates certain rules. In this section, we consider rules based on the number of IPs associated with a given token: if a token is associated with many IPs, it is tagged as suspicious.\footnote{The rules will not be exposed in detail in this article to preserve the confidentiality of the company where the study was carried out. } 
Assuming that the classification has already taken place, the operator can analyze the data related to cases classified as suspicious to understand whether the rules applied are too permissive or too rigid.

As a starting point, we consider statistics related to the detection of suspicious users. 
Taking as an example the week from 06/08/2024 to 06/15/2024, we characterize the quantiles of the distribution of times that each suspicious user was tagged as such. The 0.64 quantile equals 1, meaning that 64\% of users identified as suspects were identified as such only once. In contrast, the 0.66, 0.8, 0.95, 0.99 and 1 quantiles equal 2, 3, 59, 663, and 1397, meaning that some users were identified as suspects hundreds or even more than one thousand times. 
%
%
The above numbers suggest that users tagged as suspicious only once may correspond to false positives, but that others are clearly anomalous.

Next, we consider a longitudinal analysis of the recurrence of suspicious users. 
Figure~\ref{fig:blocks-by-day}(a) shows the number of tokens tagged as suspicious, and Figure~\ref{fig:blocks-by-day}(b) shows the subset of those tokens that are non-recurring. Note that the y axis in Figure~\ref{fig:blocks-by-day}(a) 
 varies in the range of thousands of tokens, whereas in Figure~\ref{fig:blocks-by-day}(b), it is in the range of hundreds. So, it is clear that the group of suspects with recurrence is responsible for the largest portion of tokens classified as suspicious.  

The above results indicate that in the considered week, recurring cases were prevalent. However, analyzing the period from October 2022 to October 2023 we identified that approximately 80\% of users that were once classified as suspect were never reclassified. 
Those users may correspond to false positives, motivating new classification strategies based on the volume of requests per user as opposed to the number of IPs associated with a given token, as further detailed next.

\begin{figure}[t]
    \centering
    \begin{tabular}{c}
\includegraphics[width=1\textwidth]{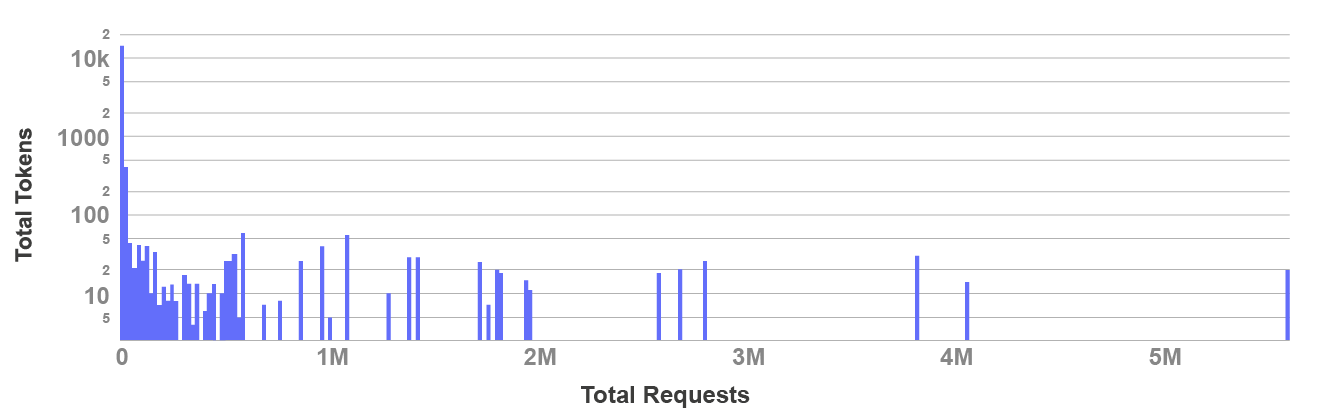}  \\
(a) \\
\includegraphics[width=1\textwidth]{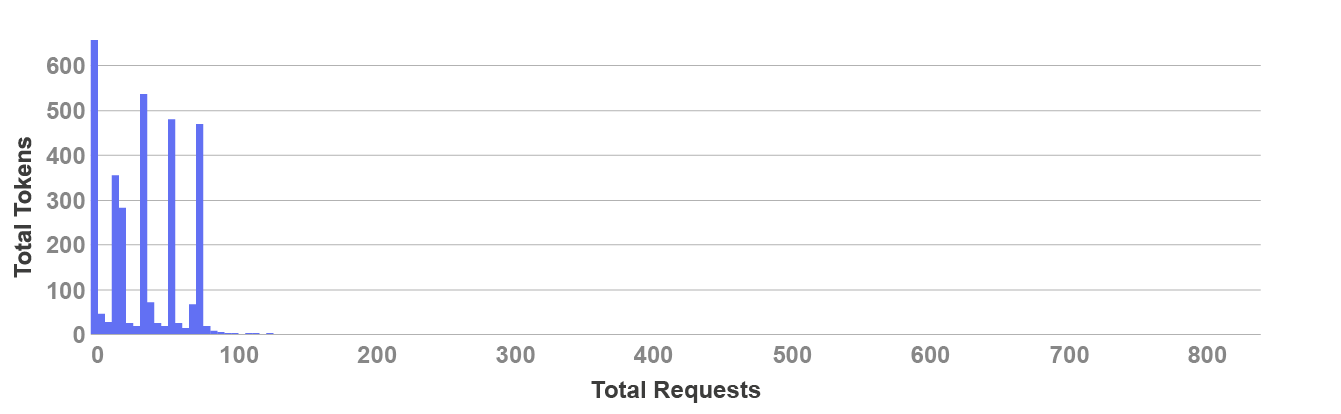} 
\\
(b) \\
\end{tabular}
    \caption{Concentration of tokens by   number of requests in (a) suspects  and (b) all data.}\label{fig:histogram-pfc} \vspace{-0.2in}
\end{figure}
\subsection{Analysis of occurrences of suspected consumption based on consumption patterns: 
volumetry of requests per user} \label{sec:volumetry2}

\begin{figure}[t]
    \centering
    \begin{tabular}{c}
\includegraphics[width=1\textwidth]{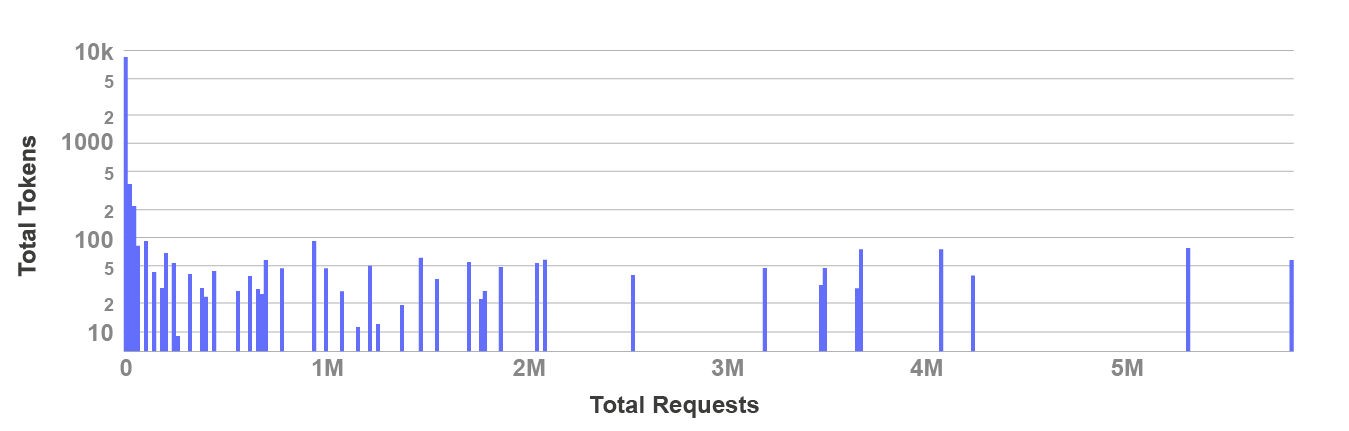}  \\
(a) \\
\includegraphics[width=1\textwidth]{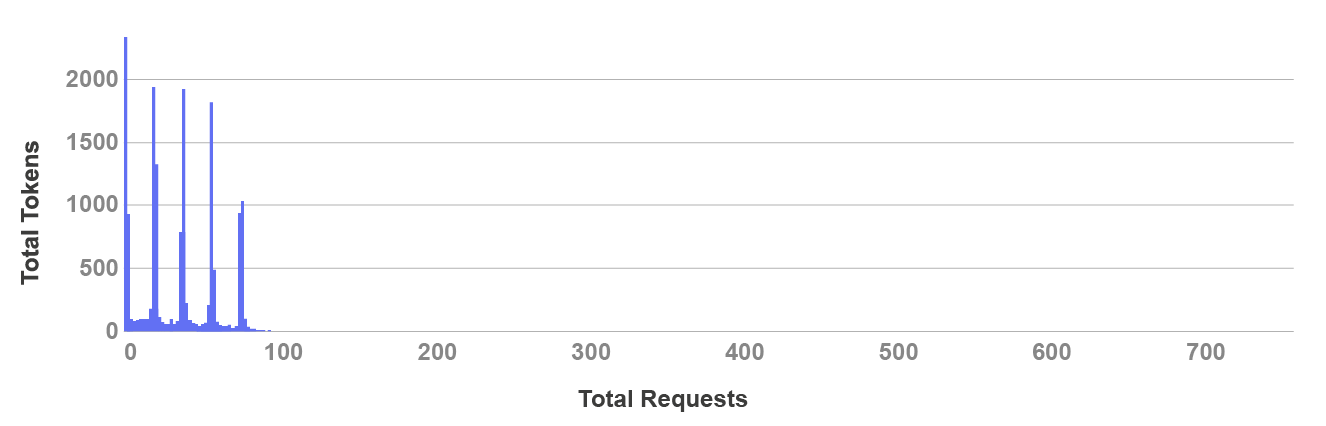}  \\
(b) 
\end{tabular}
    \caption{Concentration of tokens by   number of requests for content with a smaller chunk size (compared to Figure~\ref{fig:histogram-pfc}) in  (a) suspicious cases and (b) all cases.}
    \label{fig:histogram-sptv}
\end{figure}


Next, we consider the distribution of requests for video chunks.  
We compare a standard sample distribution with the cases that are flagged as suspects. The data is sampled from 1 out of every 100 users, regardless of whether they have been flagged as suspicious or not, within a specified time window.

Taking the channel with the highest audience as an example, as shown in Figure~\ref{fig:histogram-pfc}(a), the majority of tokens made below 1M requests, i.e., the mode of the number of requests per token is less than 1M. However, there is a group of tokens that significantly deviates from this threshold. 
In particular, the tokens with many more requests could be from malicious users or problematic connections, in which multiple attempts exist to access the content. 
Both cases deserve attention, as they may signal abuse of tokens or a network connection problem. Distinguishing between the two cases is key for determining when and if to block users.  

Whereas Figure~\ref{fig:histogram-pfc}(a) shows all requests marked as suspicious, 
 Figure~\ref{fig:histogram-pfc}(b) shows a sample from all the data, irrespective of its tag. In this sample, we typically observe a much smaller number of requests per token. This indicates that the number of requests from suspect cases is higher than usual, suggesting that analyzing the request pattern per token can aid in identifying suspicious cases.

Each content has a different consumption pattern. 
This is illustrated, for example, through the analysis of an alternative content that, in particular, has a smaller chunk size than the one analyzed in the previous figure. By observing this additional content, we notice that, as in the previous case, Figures~\ref{fig:histogram-sptv}(a) and Figures~\ref{fig:histogram-sptv}(b) exhibit very different patterns, allowing us to distinguish between typical and anomalous users. However, given the specificity of this new content, which has a smaller chunk size, it is expected that the typical number of requests will differ from that necessary to consume the previous content. Indeed, comparing 
 Figures~\ref{fig:histogram-pfc} and~\ref{fig:histogram-sptv} we note that the typical values of total requests per token vary at different ranges. 
 Those results indicate that one needs to analyze the specific consumption patterns per content, e.g., accounting for chunk sizes, to distinguish anomalous traffic.

\section{Conclusion}
\label{sec:conclusion}

This work presented two contributions to the problem of token sharing in a CDN. First, queuing models were proposed for performance analysis, which aids in understanding various piracy scenarios. Second, the analysis of real access management strategies showed that simple policies are effective in detecting inappropriate consumption, 
identifying that a significant fraction of anomalous users are recurring Future work involves exploring the use of machine learning for detecting suspicious cases and applying the proposed queuing models in real scenarios.

\section*{Acknowledgement}

 This project was sponsored by CAPES, CNPq, and FAPERJ   (315110/2020-1, E-26/211.144/2019, and E-26/201.376/2021).

%
%
%
\bibliographystyle{splncs04}
\bibliography{biblio}
\end{document}